\documentclass[aps,prl,reprint,superscriptaddress]{revtex4-1}

\usepackage{graphicx}
\usepackage{dcolumn}
\usepackage{natbib}
\usepackage{color}
\usepackage{amssymb,amsmath}
\usepackage{hyperref}
\usepackage{lipsum}

\begin{document}
	
	\title{Remarks on the Fundamentals of Time \\ \emph{oder die Zeitdämmerung}}
	
	\author{J. S. Finberg}
	\email{jsf2178@columbia.edu}
	\affiliation{Columbia University Department of Physics, New York, NY, USA}
	
	\date{\today}
	
	\begin{abstract}
	This paper argues that the common identification of time with entropy is a category error. Entropy accounts for the arrow of time but not for time itself. Even in a maximally entropic universe—the so-called heat death—temporal structure persists, grounded in the correlations of the quantum vacuum and the geometry of spacetime. The twilight of time is not its disappearance but its loss of directionality.
	\end{abstract}
	
	\maketitle
	
	%%%%%%%%%%%%%%%%%%%%%%%%%%%%%%%%%%%%%%%%%%%%%%%%%%%%%%%%%%%%%%%%%%%%%%
	%% I. Introduction
	%%%%%%%%%%%%%%%%%%%%%%%%%%%%%%%%%%%%%%%%%%%%%%%%%%%%%%%%%%%%%%%%%%%%%%
	
\section{die Zeitendämmerung}

1. Time is not defined by physics; it is assumed in order that physics may define anything else.\cite{boltzmann1964} Physics presupposes time to formulate its laws. Without time, no equation of motion can be written. The derivative $\frac{dx}{dt}$ requires $t$ as given. Velocity, acceleration, force—all require temporal ratios. These ratios presuppose what they do not explain. Without time, no measurement can be taken. To measure is to compare states. States must be ordered to be compared. Ordering is already temporal. Without time, no change can be described. Change is difference across time. Without the dimension of difference, change collapses. Physics thus requires time before it begins.

2. Newton treated time as absolute.\cite{newton1687} Absolute time flows uniformly for all observers. This flow is independent of events. It is independent of matter. It is the stage upon which physics unfolds. This time is the universal measure of change. All clocks, in principle, can be synchronized to it. All motion is ultimately referred to it. It is the hidden parameter in every Newtonian equation. Newton's time is metaphysical: presupposed, not derived. It cannot be detected directly, only through change. Yet it is posited as more fundamental than change. This priority of time over change defines classical mechanics. Leibniz objected: time without change is meaningless.\cite{leibniz1715} For Leibniz, time is the order of succession. Remove events, and time vanishes with them. Newton's reply: the mathematics requires absolute time.

3. Einstein relativized time without explaining it.\cite{einstein1916} Proper time replaces absolute time. Each worldline carries its own temporal measure. Observers disagree on simultaneity. Time becomes path-dependent. The Lorentzian manifold requires a temporal dimension. The metric signature (-,+,+,+) distinguishes time from space. Timelike intervals preserve causality. Without this distinction, relativity collapses. Relativity transforms time but still presupposes it. To speak of proper time requires time already given. The manifold exists before worldlines are traced. Minkowski: "Space and time separately have vanished."\cite{minkowski1908} Yet spacetime retains temporal structure. The unity conceals but does not eliminate the distinction.

4. Quantum mechanics externalizes time.\cite{wald1994} Time is a parameter, not an observable. Position has an operator: $\hat{x}$. Momentum has an operator: $\hat{p}$. Time has no operator: only $t$. Pauli: no time operator can exist with bounded Hamiltonian.\cite{pauli1933} If time were observable, energy spectra would be unbounded below. Physical systems would be unstable. Matter itself could not persist. Matter's stability requires time outside the operator algebra. Time frames quantum evolution. It is not subject to that evolution. The framework cannot be within what it frames.

5. Quine: to be is to be the value of a bound variable.\cite{quine1960} Physics quantifies over times: $t_1, t_2, t_3$. These appear in integrals: $\int_{t_1}^{t_2} L \, dt$. They index states: $|\psi(t)\rangle$. They order measurements. Therefore times are real. To deny their reality while using them is incoherent. Our best theories commit us to temporal ontology. This commitment is forced, not chosen. But indispensability is not explanation. We use what we cannot derive. Time is primitive in our theoretical structure. To be primitive is to be unexplained.

6. Time is presupposed across all physical theories. Presupposition entails commitment to existence. We cannot do physics without time. Therefore time exists. But this existence is assumed, not proven. Existence alone does not yield directionality. The manifold permits both orientations. The equations run both ways. Something more is needed for the arrow.

7. The laws of physics contain no distinction between past and future.\cite{price1996} Classical mechanics is time-reversible.\cite{boltzmann1964} Every trajectory can be traced backward. Reverse all velocities, and the system retraces its path. The equations are satisfied equally in both directions. Newton's $F = ma$ contains no arrow. The equations are valid under time reversal. Replace $t$ with $-t$. The physics remains unchanged. Only our interpretation differs. Laplace's demon sees no distinction. Given complete information, past and future are equally determined. The asymmetry is epistemic, not ontological. We remember the past but not the future—the equations do not.

8. Relativity treats $\pm t$ symmetrically.\cite{einstein1916} The manifold carries no inherent arrow.\cite{hawking1973} The metric is static. All events coexist tenselessly. McTaggart's B-series, not A-series.\cite{mctaggart1908} Events exist tenselessly in the block universe. Past, present, and future are perspectival labels. No moment is ontologically privileged. "Now" is indexical, like "here." The Lorentz group preserves this symmetry. Both future and past light cones are equally real. Causality restricts influence, not existence. The block simply is.

9. Quantum mechanics preserves this symmetry.\cite{wald1994} The Schrödinger equation is invariant under $t \to -t$. Unitary evolution is reversible. Information is preserved. The past could, in principle, be reconstructed. CPT symmetry holds universally. Charge, parity, and time reversal together leave physics unchanged. This is a fundamental theorem of quantum field theory. Even particle creation respects this symmetry. Feynman: antiparticles are particles moving backward in time. The mathematics supports this interpretation. Time's arrow is not in the quantum formalism. It must enter from elsewhere.

10. Experience contradicts this symmetry. Memory accumulates in one direction. We remember yesterday, not tomorrow. Records exist of the past, not the future. Information flows from past to future. Causation flows from past to future. Causes precede effects. We can influence the future, not the past. Agency itself presupposes an arrow. Prior: tense cannot be eliminated.\cite{prior1967} "Thank goodness that's over" has no tenseless translation. The lived present is irreducible. Physics describes time from outside; we experience it from within.

11. The resolution: distinguish time's existence from its direction. Existence is geometric. The manifold provides temporal structure. This structure is symmetric. It exists independently of arrow. Direction is entropic. Entropy breaks the symmetry. It selects a preferred direction. This selection is statistical, not fundamental. The laws permit both; boundary conditions select one. The universe's initial state was special. This specialness propagates forward. Hence the arrow.

12. Entropy explains the arrow, not time itself.\cite{price1996} Boltzmann: systems evolve toward higher entropy.\cite{boltzmann1964} High-entropy states vastly outnumber low-entropy states. Phase space volume counts microstates. Macrostates with more microstates are more probable. Probability drives evolution. Evolution toward disorder is statistical, not dynamical. Individual trajectories are reversible. Ensembles show irreversibility. The arrow emerges from counting. Loschmidt's paradox: why this direction?\cite{loschmidt1876} The dynamics are symmetric. Statistical mechanics assumes molecular chaos. This assumption smuggles in the arrow.

13. The arrow requires a special initial condition.\cite{albert2000} The Past Hypothesis: the universe began in low entropy. The Big Bang was extraordinarily ordered. This order has been dissipating since. All arrows trace back to this origin. Without this, no universal arrow exists. Generic initial conditions yield no arrow. Entropy would increase in both temporal directions. The Past Hypothesis breaks the symmetry. The arrow is contingent, not necessary. Another universe might have different boundary conditions. Or high entropy at both ends. The arrow is a fact about our universe, not about time itself.

14. Entropy presupposes time. To write $S(t_2) > S(t_1)$ requires temporal ordering. The relation "greater than" applies to entropy values. These values are indexed by time. Remove time, and the comparison becomes meaningless. The gradient requires the dimension. Entropy is a function: $S: t \to \mathbb{R}$. Functions require domains. Time is entropy's domain. To identify time with entropy is a category error. It confuses the function with its domain. It mistakes the map for the territory. Time is the stage; entropy is the play.

15. Therefore: entropy gives time an arrow, not existence. Time without entropy would lack direction. But it would still exist as pure succession. The manifold would persist. Only orientation would vanish. Entropy without time is incoherent. "Increase" is a temporal concept. Static entropy is not an arrow. The arrow requires both entropy and time.

16. The vacuum sustains temporal structure.\cite{wald1994} The vacuum is not empty but structured. Zero-point fluctuations persist irreducibly. Every mode has energy $E_0 = \frac{1}{2}\hbar\omega$. This energy cannot be removed. The ground state still oscillates. The uncertainty principle forbids absolute stillness. $\Delta x \Delta p \geq \frac{\hbar}{2}$. Perfect position implies infinite momentum uncertainty. The vacuum must fluctuate. Virtual particles continually appear and annihilate. Electron-positron pairs flicker in and out. The vacuum is a foam of virtual processes. These processes are temporal.

17. These fluctuations encode temporality. Correlation functions decay with time separation. $\langle \phi(x,t) \phi(x,0) \rangle$ depends on $t$. This dependence defines temporal structure. Even in vacuum, time leaves traces. The Casimir effect demonstrates vacuum activity. Parallel plates attract in vacuum. The force comes from vacuum fluctuations. Fluctuations imply temporal process. Fluctuations provide continual "ticks." Not mechanical ticks of a clock. But structural oscillations of fields. The vacuum itself is a clock.

18. Heat death does not erase vacuum structure. Entropy maximization eliminates gradients. No more stars, no more life. No more information processing. Maximum disorder. But quantum uncertainty remains. The uncertainty principle is not thermodynamic. It is fundamental to quantum mechanics. It survives equilibrium. The vacuum persists after thermal equilibrium. Fields remain. Fluctuations continue. Time endures.

19. Therefore vacuum grounds time's persistence. Geometry provides structure. But geometry alone is static. It requires animation. The vacuum provides this animation. Matter provides clocks. But matter eventually decays. Protons disintegrate. Only vacuum remains. Vacuum is time's ultimate foundation. More fundamental than matter. More persistent than entropy gradients. The substrate of temporality itself.

20. Even at maximum entropy, time endures.\cite{carroll2010} De Sitter space has a horizon temperature.\cite{gibbons1977} $T = \frac{\hbar H}{2\pi k_B}$ even in vacuum. This is the Gibbons-Hawking temperature. It arises from the cosmological horizon. The horizon radiates thermally. Thermal radiation defines temporal intervals. Particles are created and annihilated. These events mark time. The vacuum is never truly cold. Detectors click at rates determined by proper time. An Unruh-DeWitt detector registers particles. Click rates measure temporal passage. Time is operationally definable even at equilibrium.

21. The thermal time hypothesis.\cite{connes1994} Time emerges from statistical states. The KMS condition defines temporal evolution. Statistical states induce dynamics. Time is not external but emergent. Correlations define temporal flow. Even without external clocks. The state itself becomes the clock. Rovelli: "Time is thermodynamical."\cite{rovelli1993} Even without matter, succession persists. The vacuum state has correlations. Correlations evolve. Evolution is time.

22. What ends is the arrow, not time. Direction ceases when entropy maximizes. No more increase possible. Statistical equilibrium achieved. The arrow vanishes. But duration continues through fluctuation. Quantum fluctuations are eternal. They provide temporal structure. Structure without direction is still structure. Time persists without asymmetry. Like space without a compass. Dimension remains when orientation is lost. The twilight of time is not darkness but symmetry.

23. Time transcends both change and geometry. Aristotle: time is the measure of change.\cite{aristotle_physics} Without motion, no time. Time counts the before and after. Stillness would be timeless. Time depends on change. But change presupposes succession. To change is to be different. Difference requires comparison. Comparison requires temporal ordering. The definition is circular. Time requires change. Change requires time. Neither can ground the other.

24. Modern physics embeds time in geometry. Spacetime provides temporal structure. The manifold exists complete. All events have coordinates. Time is geometrized. But geometry alone is static. The block universe does not flow. It simply is. Geometry provides structure, not dynamics. Animation requires more than structure. The manifold must be traversed. Worldlines must be traced. Something must move through the geometry.

25. The Wheeler-DeWitt equation.\cite{dewitt1967} $H\Psi = 0$ suggests timelessness. The universe as a whole has no external time. The wave function does not evolve. Time disappears. But subsystems evolve relationally. One subsystem serves as clock for another. Internal time emerges. Page-Wootters mechanism.\cite{page1983} Time emerges from correlations. Entanglement defines temporal relations. No external parameter needed. Time is relational, not absolute.

26. Time is irreducible. Not reducible to change alone. The vacuum fluctuates without macroscopic change. Time persists in the fluctuations. Change is sufficient but not necessary. Not reducible to geometry alone. Geometry without dynamics is frozen. The block requires traversal. Structure is necessary but not sufficient. Not reducible to entropy alone. Entropy provides arrow, not existence. Maximum entropy does not eliminate time. Direction is contingent; time is necessary.

27. Time is the synthesis. Geometry provides possibility. The framework for events. The stage for physics. Structure without content. Change provides actuality. Events occur. Things happen. Content within structure. Vacuum provides persistence. When change ceases, fluctuations remain. When matter decays, fields persist. The ultimate substrate endures.

28. This trinity ensures time's permanence. One pillar may fall. Change may cease at heat death. But geometry and vacuum remain. Time survives. Two pillars may fall. In principle, only vacuum might remain. Still time persists in fluctuations. The last pillar holds. All three cannot fall while the universe exists. To exist is to be in time. The universe cannot escape temporality. Time is the condition of existence itself.

29. What we call time's death is only its loss of direction. The arrow fades; the axis remains. Entropy ceases to increase; succession continues. The universe persists in time without time's arrow. This is the twilight: not darkness but the dimming of distinction. Past and future become indistinguishable, not nonexistent. Time enters its final symmetry, purged of preference, but not of presence.

30. The twilight of time is not its vanishing, but its paling into symmetry.

\section{Acknowledgments}
	
	\begin{acknowledgments}
	The author also thanks E.C.H. for her support even though she often has no idea what this is even about. Love you babe! 
	\end{acknowledgments}
	
	\bibliographystyle{apsrev4-1}
	\bibliography{Refs}
	\nocite{*}
	
\end{document}